\documentclass[prd,superscriptaddress,nofootinbib,amsmath,amssymb,aps,11pt]{revtex4}

\usepackage{bm}
\usepackage{amsfonts}
\usepackage{latexsym}
\usepackage[latin1]{inputenc}
\usepackage{graphicx}
\usepackage{amsmath}
\usepackage{palatino}
\usepackage{mathpazo}
\usepackage[outdir=./]{epstopdf}

\usepackage{booktabs}
\usepackage{dcolumn}

\def\jnl@style{\it}
\def\aaref@jnl#1{{\jnl@style#1}}

\def\aaref@jnl#1{{\jnl@style#1}}

\def\aj{\aaref@jnl{AJ}}                   
\def\apj{\aaref@jnl{ApJ}}                 
\def\apjl{\aaref@jnl{ApJ}}                
\def\apjs{\aaref@jnl{ApJS}}               
\def\apss{\aaref@jnl{Ap\&SS}}             
\def\aap{\aaref@jnl{A\&A}}                
\def\aapr{\aaref@jnl{A\&A~Rev.}}          
\def\aaps{\aaref@jnl{A\&AS}}              
\def\mnras{\aaref@jnl{Mon.~Not.~Roy.~Astron.~Soc.}}             
\def\prd{\aaref@jnl{Phys.~Rev.~D}}        
\def\prc{\aaref@jnl{Phys.~Rev.~C}}  
\def\prl{\aaref@jnl{Phys.~Rev.~Lett.}}    
\def\qjras{\aaref@jnl{QJRAS}}             
\def\skytel{\aaref@jnl{S\&T}}             
\def\ssr{\aaref@jnl{Space~Sci.~Rev.}}     
\def\zap{\aaref@jnl{ZAp}}                 
\def\nat{\aaref@jnl{Nature}}              
\def\aplett{\aaref@jnl{Astrophys.~Lett.}} 
\def\apspr{\aaref@jnl{Astrophys.~Space~Phys.~Res.}} 
\def\physrep{\aaref@jnl{Phys.~Rep.}}      
\def\physscr{\aaref@jnl{Phys.~Scr}}       
\def\commat{\aaref@jnl{Comm.~Math.~Phys.}}              
\def\science{\aaref@jnl{Science}}               
\def\cqg{\aaref@jnl{Classical Quant.~Grav.}}            
\def\jpcs{\aaref@jnl{JPCS}}                                     
\def\ijmpd{\aaref@jnl{Int.~J.~Mod.~Phys.~D}}                    
\def\grg{\aaref@jnl{Gen.~Relat.~Gravit.}}               
\def\rpp{\aaref@jnl{Rep.~Prog.~Phys.}}          
\def\npa{\aaref@jnl{Nucl.~Phys.~A}}        
\def\lrr{\aaref@jnl{Living Rev.~Rel.}}                   
\def\jcap{\aaref@jnl{J.~Cosmology Astropart.~Phys.}}    
\def\rmp{\aaref@jnl{Rev.~Mod.~Phys.}}   

\allowdisplaybreaks[1]

\begin{document}

\title{Mixed configurations of tensor-multi-scalar solitons and neutron stars}

\author{Daniela D. Doneva}
\email{daniela.doneva@uni-tuebingen.de}
\affiliation{Theoretical Astrophysics, Eberhard Karls University of T\"ubingen, T\"ubingen 72076, Germany}
\affiliation{INRNE - Bulgarian Academy of Sciences, 1784  Sofia, Bulgaria}

\author{Stoytcho S. Yazadjiev}
\email{yazad@phys.uni-sofia.bg}
\affiliation{Theoretical Astrophysics, Eberhard Karls University of T\"ubingen, T\"ubingen 72076, Germany}
\affiliation{Department of Theoretical Physics, Faculty of Physics, Sofia University, Sofia 1164, Bulgaria}
\affiliation{Institute of Mathematics and Informatics, 	Bulgarian Academy of Sciences, 	Acad. G. Bonchev St. 8, Sofia 1113, Bulgaria}


\begin{abstract}
In the present paper we consider special classes of massive  tensor-multi-scalar theories of gravity whose target space metric admits
Killing field(s) with a periodic flow. For such  tensor-multi-scalar theories we show that there exist
mixed configurations of tenor-multi-scalar solitons and relativistic (neutron) stars. The influence of the curvature of the 
target space on the structure of the mixed configurations is studied for two explicit scalar-tensor theories.  
The stability of the obtained compact objects is examined and it turns out that the stability region spans to larger central energy densities and central values of the scalar field compared to the pure neutron stars or the pure tensor-multi-scalar solitons. 
    
\end{abstract}

\maketitle

\section{Introduction}

Ones of the viable alternative theories of gravity that pass through all the observations and measurements so far are the tensor-multi-scalar
theories \cite{Damour_1992,Horbatsch_2015}. In these theories the gravitational interaction is mediated not only by the spacetime metric but also by additional scalar fields. Special classes of the tensor-multi-scalar theories are those  whose target space metric admits Killing field(s) with a periodic flow. For such tensor-multi-scalar theories we showed \cite{Yazadjiev_2019} that if the dynamics
of the scalar fields is confined on the periodic orbits of the Killing field(s) then there exist new solutions
describing dark compact objects -- the tensor-multi-scalar solitons formed by a condensation of the gravitational
scalars.  Since there is no direct interaction between the gravitational scalars and the electromagnetic field
the tensor-multi-scalar solitons are  indeed dark in nature. In agreement with the present day observations their mass can range at least from the mass of a neutron star  to the mass of dark objects in the center of the galaxies (and even more) in dependence  of mass(es) of the gravitational scalars and which massive sector is excited. These facts show that the tensor-multi-scalar solitons could have  important implications for the dark matter problem.  The existence of the tensor-multi-scalar solitons  points towards the possibility that part of dark matter is made of  condensed gravitational scalars.  As a matter of fact these object in their most simplified version coincide with the well know boson stars \cite{Schunck2008} but the tensor-multi-scalar solitons posses much richer phenomenology as described in detail in \cite{Yazadjiev_2019}.

In the present paper we extend further our study of the tensor-multi-scalar solitons and show that there exist mixed configurations of 
tensor-multi-scalar solitons and relativistic (neutron) stars when the mass of the excited gravitational scalars are in certain range. 
In Section II we discuss the main theoretical background behind the considered class of tensor-multi-scalar theories and mixed compact objects. The numerical solutions and their stability is discussed in Section III. The paper ends with discussion.

\section{Tensor-multi-scalar solitons and mixed configurations -- analytical background} 

Within the framework of  tensor-multi-scalar theories of gravity \cite{Damour_1992,Horbatsch_2015} the gravitational interaction is mediated by the spacetime metric $g_{\mu\nu}$ and $N$ scalar fields $\varphi^{a}$  which take value in a coordinate patch of an N-dimensional Riemannian (target) manifold ${\cal E}_{N}$ supplemented with positively definite metric $\gamma_{ab}(\varphi)$. In the Einstein frame the action of the  tensor-multi-scalar theories of gravity is given by

\begin{eqnarray}\label{Action}
S= \frac{1}{16\pi G_{*}}\int d^4\sqrt{-g}\left[R - 2g^{\mu\nu}\gamma_{ab}(\varphi)\nabla_{\mu}\varphi^{a}\nabla_{\nu}\varphi^{b} - 4V(\varphi)\right]  + S_{matter}(A^{2}(\varphi) g_{\mu\nu}, \Psi_{matter}).
\end{eqnarray}
Here $G_{*}$ is the bare gravitational constant, $\nabla_{\mu}$ and $R$ are the covariant derivative  and the Ricci scalar curvature with respect to  the Einstein frame metric $g_{\mu\nu}$, and $V(\varphi)\ge 0$ is the potential of the scalar fields. In order for the weak equivalence principle to be satisfied the matter fields, denoted collectively by $\Psi_{matter}$, are coupled only to the physical Jordan metric ${\tilde g}_{\mu\nu}= A^2(\varphi) g_{\mu\nu}$ where  $A^2(\varphi)$ is the conformal factor relating the Einstein and the Jordan metrics, and which, together with $\gamma_{ab}(\varphi)$ and $V(\varphi)$, specifies the tensor-multi-scalar theory.  

Varying  action (\ref{Action}) with respect to the Einstein frame metric $g_{\mu\nu}$ and the scalar fields $\varphi^a$ we obtain the Einstein frame field equations

\begin{eqnarray}\label{FE}
&&R_{\mu\nu}= 2\gamma_{ab}(\varphi) \nabla_{\mu}\varphi^a\nabla_{\nu}\varphi^b + 2V(\varphi)g_{\mu\nu} + 8\pi G_{*} \left(T_{\mu\nu} - \frac{1}{2}T g_{\mu\nu}\right), \\
&&\nabla_{\mu}\nabla^{\mu}\varphi^a = - \gamma^{a}_{\, bc}(\varphi)g^{\mu\nu}\nabla_{\mu}\varphi^b\nabla_{\nu}\varphi^c 
+ \gamma^{ab}(\varphi) \frac{\partial V(\varphi)}{\partial\varphi^{b}} - 
4\pi G_{*}\gamma^{ab}(\varphi)\frac{\partial\ln A(\varphi)}{\partial\varphi^{b}}T, \nonumber
\end{eqnarray}
with $T_{\mu\nu}$ being  the Einstein frame energy-momentum tensor of matter and $\gamma^{a}_{\, bc}(\varphi)$ being 
the Christoffel symbols with respect to the target space metric $\gamma_{ab}(\varphi)$. From the field equations and the contracted Bianchi identities we also find the following conservation law for the Einstein frame energy-momentum tensor

\begin{eqnarray}\label{Bianchi}
\nabla_{\mu}T^{\mu}_{\nu}= \frac{\partial \ln A(\varphi)}{\partial \varphi^{a}}T\nabla_{\nu}\varphi^a .
\end{eqnarray}

The Einstein frame energy-momentum tensor $T_{\mu\nu}$ and the Jordan frame one ${\tilde T}_{\mu\nu}$
are related via the formula $T_{\mu\nu}=A^{2}(\varphi){\tilde T}_{\mu\nu}$. As usual, in the present paper  the matter content of 
the stars will be described as a perfect fluid. In the case of a perfect fluid the relations between the energy density,
pressure and 4-velocity in both frames are given by $\rho=A^{4}(\varphi){\tilde \rho}$, $p=A^{4}(\varphi){\tilde p}$ and 
$u_{\mu}=A^{-1}(\varphi) {\tilde u}_{\mu}$. 
         
Following our previous paper \cite{Yazadjiev_2019} we shall consider a special class of tensor-multi-scalar theories which is defined as follows. We require that the metric $\gamma_{ab}(\varphi)$ admits a Killing field  $K^{a}$ with  a periodic flow and also, $A(\varphi)$ and $V(\varphi)$  be invariant under the flow of the Killing field $K^{a}$, i.e. ${\cal L}_{K} A(\varphi)=K^a\partial_{a}A(\varphi)=0$  and ${\cal L}_{K}V(\varphi)= K^a\partial_{a}V(\varphi)=0$.  As shown in \cite{Yazadjiev_2019} the existence of a  Killing field $K^a$ gives rise to the following Jordan frame conserved current ${\tilde J}^{\mu}$ given by  

\begin{eqnarray}\label{JFC}
{\tilde J}^{\mu}= \frac{1}{4\pi G(\varphi)} {\tilde g}^{\mu\nu} K_{a}\partial_{\nu}\varphi^{a}
\end{eqnarray}
where $G(\varphi)= G_{*}A^2(\varphi)$. The conserved current  exists even in the presence of matter when our requirements are satisfied. This can be proven by using  the fact that $K^a$  is a Killing field for $\gamma_{ab}(\varphi)$, the  field equations (\ref{FE}) and  ${\cal L}_K A(\varphi)={\cal L}_{K}V(\varphi)=0$.

In order to avoid unnecessary technical complications we shall focus on the simplest  $N=2$ case. Nevertheless our approach presented below is general and, leaving aside  some technical details, is applicable to the case with arbitrary $N$. 
Following \cite{Yazadjiev_2019} in order to be more specific, we shall focus on maximally symmetric ${\cal E}_2$. In this case the metric $\gamma_{ab}(\varphi)$
presented in the global isothermal (or conformally flat) coordinates is given by 
\begin{eqnarray}
\gamma_{ab}(\varphi)= \Omega^2(\psi)\delta_{ab},
\end{eqnarray} 
where $\psi^2=\delta_{ab}\varphi^a\varphi^b$ and the explicit form of the conformal factor is 

 \begin{eqnarray}
\Omega^2= \frac{1}{\left(1+ \frac{\kappa}{4}\delta_{ab}\varphi^a\varphi^b\right)^2} =\frac{1}{\left(1+ \frac{\kappa}{4}\psi^2\right)^2}
\end{eqnarray} 
where the constant $\kappa$ is the  curvature of ${\cal E}_2$.    

In these coordinates the Killing field with the periodic orbits is explicitly given by 
\begin{eqnarray}
K=\varphi^{2}\frac{\partial}{\partial\varphi^1} - \varphi^{1}\frac{\partial}{\partial\varphi^2}. 
\end{eqnarray}

In order for  the conditions ${\cal L}_K A(\varphi)=0$ and ${\cal L}_{K}V(\varphi)=0$  to be satisfied $A(\varphi)$ and $V(\varphi)$ must depend on $\varphi^a$ through  $\psi$, i.e. $A=A(\psi)$ and $V=V(\psi)$. 

In accordance with the main purpose of the present paper we consider strictly static and spherically symmetric asymptotically flat spacetime.
The everywhere timelike Killing field will be denoted by $\xi$ and in adapted coordinates can be written in the standard form $\xi=\frac{\partial}{\partial t}$. The spacetime metric takes the usual form

\begin{eqnarray}
ds^2= - e^{2\Phi(r)}dt^2 + e^{2\Lambda(r)}dr^2 + r^2(d\theta^2 + \sin^2\theta d\phi^2).
\end{eqnarray} 
 
The staticity conditions that have to be imposed on the perfect fluid are well-known. That is why following \cite{Yazadjiev_2019}  we shall comment only on the conditions on the scalar fields and their effective energy-momentum tensor in order to have static geometry and static perfect fluid.  
In order for the solitons to exists the  dynamics of the scalar fields is ``confined'' on the periodic orbits of the Killing field $K^a$. In geometrical terms this can be expressed in the following way 

\begin{eqnarray}\label{periodic}
{\cal L}_{\xi}\varphi^a= -\omega K^a 	
\end{eqnarray}	
where $\omega$ is a nonzero (real) constant.  With this condition imposed  it is not difficult to check that the effective energy-momentum tensor $T^{(\varphi)}_{\mu\nu}=(4\pi G_{*})^{-1}\left[\gamma_{ab}(\varphi) (\nabla_{\mu}\varphi^a\nabla_{\nu}\varphi^b -\frac{1}{2}
g_{\mu\nu}\nabla_{\sigma}\varphi^a\nabla^{\sigma}\varphi^b) - V(\varphi)g_{\mu\nu}\right]$ of the gravitational scalars is static, i.e. ${\cal L}_{\xi}T^{(\varphi)}_{\mu\nu}=0$. One more consistency condition that must be satisfied  is the Ricci staticity condition $R[\xi]\wedge \hat{\xi}=0$ where $R[\xi]=\xi^{\mu}R_{\mu\nu}dx^\nu$ is the Ricci one-form and $\hat{\xi}=\xi_\mu dx^\mu$ is the Killing one-form naturally corresponding to the Killing field $\xi$. In view of (\ref{periodic}) the Ricci staticity condition reduces to 
\begin{eqnarray}\label{staticity}
{\hat J}\wedge {\hat \xi} = 0
\end{eqnarray}
with ${\hat J}=J_{\mu}dx^\mu$. We have to mention that our requirements automatically ensure that
the Jordan frame metric ${\tilde g}_{\mu\nu}=A^2(\varphi) g_{\mu\nu}$ is also static. This follows from the fact that  the conformal factor $A(\varphi)$ is  static -- indeed we have ${\cal L}_{\xi} A(\varphi)= \partial_a A(\varphi){\cal L}_{\xi}\varphi^a=-\omega K^a \partial_{a}A(\varphi)=0$.

In general not all solutions of (\ref{periodic}) satisfy (\ref{staticity}).
A physically natural solution to (\ref{periodic}), for which the condition (\ref{staticity}) is also satisfied, is given by 

\begin{eqnarray}
(\varphi^1, \varphi^2) = (\psi(r) cos(\omega t), \psi(r) \sin(\omega t)).
\end{eqnarray}

Taking into account all the above constructions, the dimensionally reduced vacuum field equations are the following 

\begin{eqnarray}\label{DRE}
&&\frac{2}{r}e^{-2\Lambda} \Lambda^\prime + \frac{1}{r^2}\left(1- e^{-2\Lambda}\right)=8\pi G_{*}A^{4}(\psi){\tilde \rho}  + \Omega^{2}(\psi)\left[\omega^2e^{-2\Phi}\psi^2 + 
(\psi^\prime)^2e^{-2\Lambda}\right] + 2V(\psi), \nonumber \\
&&\frac{2}{r}e^{-2\Lambda} \Phi^\prime - \frac{1}{r^2}\left(1- e^{-2\Lambda}\right)= 8\pi G_{*}A^{4}(\psi){\tilde p}+ \Omega^{2}(\psi)\left[\omega^2e^{-2\Phi}\psi^2 + 
(\psi^\prime)^2e^{-2\Lambda}\right] - 2V(\psi),\\
&&\psi^{\prime\prime} + \left(\Phi^\prime - \Lambda^\prime  + \frac{2}{r}\right)\psi^\prime + 
\left[ \omega^2e^{-2\Phi}\left(1 + 2\frac{\partial\ln\Omega}{\partial\psi^2}\psi^2\right) + 2\frac{\partial\ln\Omega}{\partial\psi^2}(\psi^\prime)^2 e^{-2\Lambda} \nonumber \right. \\ 
&&\left. 
 - 2\Omega^{-2}(\psi) \frac{\partial V(\psi)}{\partial\psi^2}\right]e^{2\Lambda}\psi - 4\pi G_{*}\Omega^{-2}(\psi) A^{4}(\psi)\frac{\partial \ln A(\psi)}{\partial\psi}({\tilde \rho} - 3{\tilde p})e^{2\Lambda}=0.
\nonumber
\end{eqnarray}

We also have to add the equation describing the static equilibrium of the perfect fluid

\begin{eqnarray}\label{HSE}
{\tilde p}^{\prime}= - ({\tilde \rho} + {\tilde p}) \left(\Phi^{\prime} + 2\frac{\partial \ln A(\psi)}{\partial \psi^2} \psi \psi^{\prime}  \right)
\end{eqnarray}
which follows from (\ref{Bianchi}).
   
The equations (\ref{DRE})--(\ref{HSE}) supplemented with the equation of state for the star matter ${\tilde p}={\tilde p}(\tilde \rho)$  and appropriate boundary conditions, describe the structure of the mixed soliton-star configurations.  The system of equations (\ref{DRE})--(\ref{HSE})
is in fact a nonlinear eigenvalue problem for $\omega$ and the natural boundary conditions are the following. The central values 
${\tilde \rho}(0)$ and $\psi(0)$, denoted by ${\tilde \rho}_c$ and $\psi_c$ below, are free parameters. The regularity at the center of the configurations requires $\Lambda(0)=1$ and $\psi^{\prime}(0)=0$. The asymptotic flatness imposes $\Phi(\infty)=\Lambda(\infty)=\psi(\infty)=0$.  In this paper we consider only 
nodeless solutions for $\psi(r)$.

As usual, the
coordinate radius $r_s$ of the star is determined by the condition ${\tilde p}(r_s)=0$ while the physical radius
of the star as measured in the physical Jordan frame is given by $R_s=A(\varphi(r_s))r_s$. 

The conserved current (\ref{JFC})  leads to a conserved charge ${\tilde q}$  given by \cite{Yazadjiev_2019} 
\begin{eqnarray}
{\tilde q}=\frac{1}{ G_{*}} \int^{+\infty}_{0} (\omega e^{-\Phi})\Omega^2(\psi)\psi^2 e^{\Lambda}r^2dr.
\end{eqnarray} 
It is however  more convenient to use the conserved charge $Q=m {\tilde q}$ instead of ${\tilde q}$. Our  numerical results will be presented namely in terms of $Q$.

Further we can define an effective radius ${\tilde R}_{sol}$ of the solitons. The Jordan frame radius of the solitons is defined as in \cite{Yazadjiev_2019}, namely   
\begin{eqnarray}
{\tilde R}_{sol}= \frac{1}{ qG_{*}} \int^{+\infty}_{0} r A(\varphi) (\omega e^{-\Phi})\Omega^2(\psi)\psi^2 e^{\Lambda}r^2dr.
\end{eqnarray}   

The other global quantities of the star in which we are interested are the mass $M$ and the  baryon rest mass $M_{0}$ of the star. The mass $M$ is defined as the Arnowitt-Deser-Misner (ADM) mass in the Einstein frame. Since the scalar fields drop of exponentially at infinity the mass $M$ also coinsides with
the ADM mass in the Jordan frame. Concerning the  baryon rest mass $M_{0}$ of the star, the definition in the physical Jordan frame is the usual
one

\begin{eqnarray}
M_{0}=\int_{Star} m_b {\tilde n}{\tilde u}^{t}\sqrt{-{\tilde g}} d^3x
\end{eqnarray}
where  ${\tilde n}$ is the baryon  number density and $m_b$ is the baryon mass. This equation, expressed in terms of
the four-velocity and metric in the Einstein frame, takes the form 

\begin{eqnarray}
M_{0}=\int_{Star} A^3(\psi)m_b {\tilde n} u^{t}\sqrt{- g} d^3x= 4\pi \int^{r_s}_{0} A^{3}(\psi) m_b {\tilde n} e^{\Lambda} r^2 dr . 
\end{eqnarray}

Finishing this section let us comment on the following. As we discussed in detail in \cite{Yazadjiev_2019}
the tensor-multi-scalar soliton can be viewed as generalization of the standard boson stars, however in many aspects they can differ considerably from them and this formal analogy with the standard boson stars is fainting down with the increasing of $N$  and the complexity of the target space metric $\gamma_{ab}$. The mixed soliton-star configurations remind us of boson-fermion stars \cite{Henriques_1990a}--\cite{Lieblig_2017} with $\psi$ being some effective boson field with an exotic kinetic  term. However, what makes the difference with the standard boson-fermion stars even drastic is the highly nonstandard coupling between the effective boson field and the matter as seen from the dimensionally reduced field equations (\ref{DRE}). It is seen that the effective boson field is sourced by the trace of the perfect fluid  energy-momentum tensor in the Einstein frame and the coupling  between $\psi$  and the perfect fluid  depends  very strongly on the particular scalar-tensor theory and can be very strong in comparison with the standard boson-fermion stars.

\section{Numerical results}

In the present paper we shall consider the simplest massive potential for the scalar field, namely 
\begin{eqnarray}
V(\varphi)= \frac{1}{2}m^2\psi^2 
\end{eqnarray}
where $m$ is the mass of the scalar fields. As in \cite{Yazadjiev_2019} we could also consider  self-interaction of the form $\lambda \psi^4$.
Since the problem  under consideration is many-parametric we prefer to focus on the principle problem without going into all possible details.
That is why in the present paper we shall neglect the self-interaction of the scalar fields.  
The influence of the self-interaction term was separately studied  for tensor-multi-scalar solitons \cite{Yazadjiev_2019} and neutron stars \cite{Staykov_2018}.

We shall focus on two massive tensor-multi-scalar theories with $A(\psi)=e^{\frac{1}{2}\beta\psi^2}$ and $A(\psi)=e^{\frac{1}{4}\gamma\psi^4}$ where $\beta$ and $\gamma$ are constants.  

The equation of state (EOS) of the baryonic mater we employed is a realistic nuclear matter EOS named APR4 \cite{APR4_EOS} and we used the piecewise polytropic approximation version of this EOS \cite{Read2009}.

\subsection{Theory with $A(\varphi)= \exp(\frac{1}{2}\beta\psi^2)$} 

This massive tensor-multi-scalar theory with $\beta<0$ is interesting because it is indistinguishable from general relativity in the weak field limit 
and differs considerably from it in the strong field regime -- the theory exhibits spontaneous scalarization for neutron stars \cite{Ramazanoglu_2016, Yazadjiev_2016}.      

Binary pulsars provide some of the tightest current constraints on scalar-tensor theories of gravity and impose strong limits on the parameter $\beta$ allowing only for very small deviations from GR \cite{Freire2012,Antoniadis2013}. These constrain are valid, though, in the case of a massless scalar field. The inclusion of mass changes the picture considerably -- the scalar field is exponentially suppressed at distances larger than its Compton wavelength and so does the scalar gravitational waves. In order to avoid the tight constraints coming from the binary pulsar observations, provided that the neutron stars are scalarized, a rough constraint on the mass $m$ can be obtained as in \cite{Ramazanoglu_2016,Yazadjiev_2016}, namely  $10^{-16}{\rm eV} <m< 10^{-9} {\rm eV}$ that translates to $10^{-6} <m< 10$ in our dimensionless units. The lower bound guaranties that the emitted scalar radiation is negligible while the upper bound guaranties that the mass term does not prevent the scalarization of the neutron star. These constraints, though, can not be directly applied in our case since we are considering multiple scalar fields and in addition we do not consider scalarization. Nevertheless, we will still consider more conservative scalar field masses in the this range. Moreover, as we will show below, the transition from baryon dominated compact object to scalar field dominated one happens exactly for masses in the interval above.

We have a three parameter family of solutions, the parameters being the mass of the scalar field $m$, the curvature constant $\kappa$ and the parameter $\beta$ in the conformal factor. In order to simplify the presentation we will fix $\beta=-6$ and present results for varying $\kappa$ and $m$. Apart from these three parameters, what determines a particular compact object solution is the central energy density of the baryonic matter ${\tilde \rho}_c$ and the central value of the scalar field $\psi_c$. Thus, a big portion of the results is presented in the form of three dimensional plots with $x$ and $y$ axis being ${\tilde \rho}_c$ and $\psi_c$ respectively.

\begin{figure}
	\includegraphics[width=0.48\textwidth]{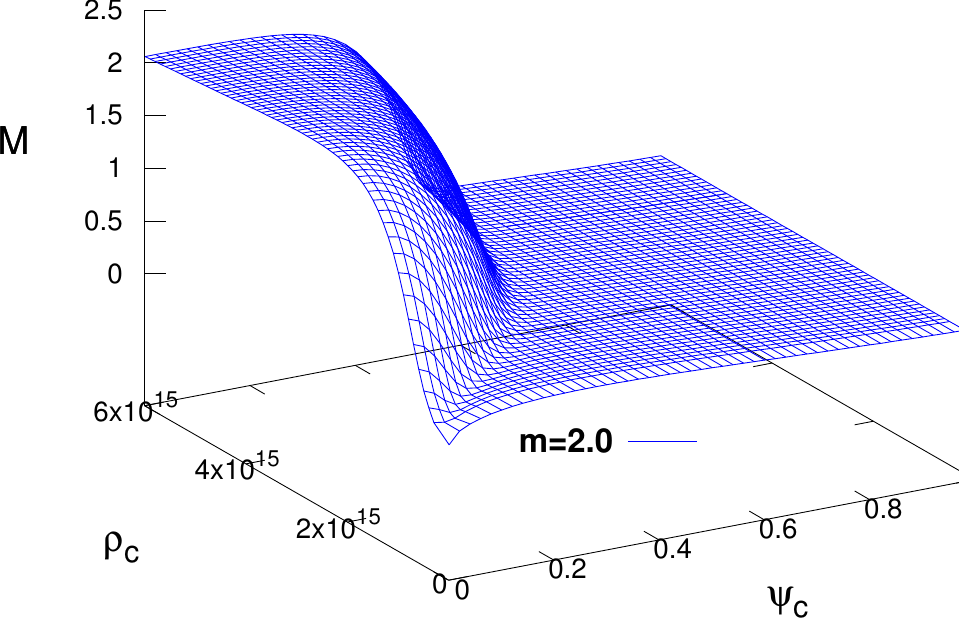}
	\includegraphics[width=0.48\textwidth]{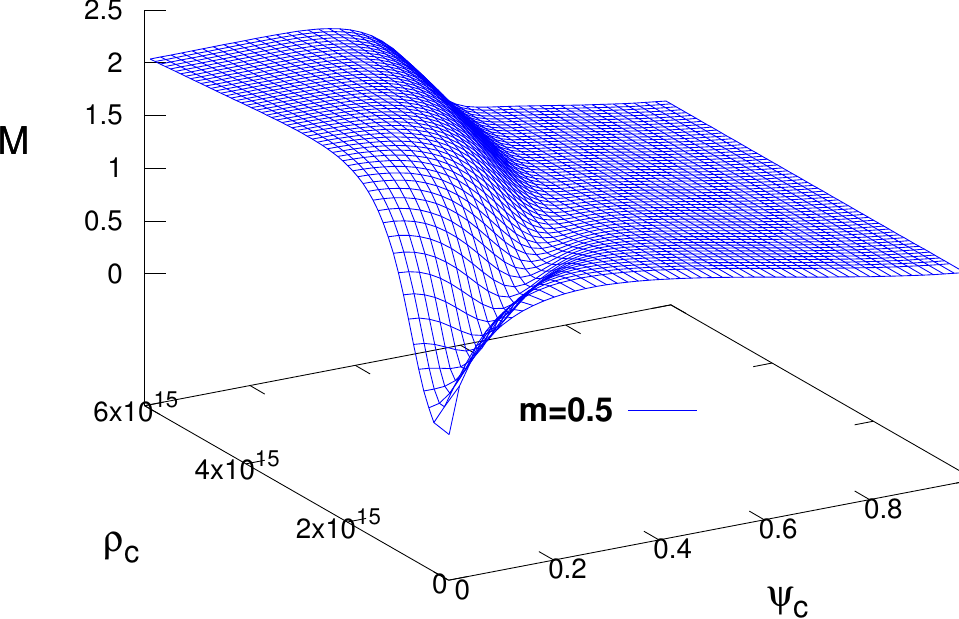}
	\includegraphics[width=0.48\textwidth]{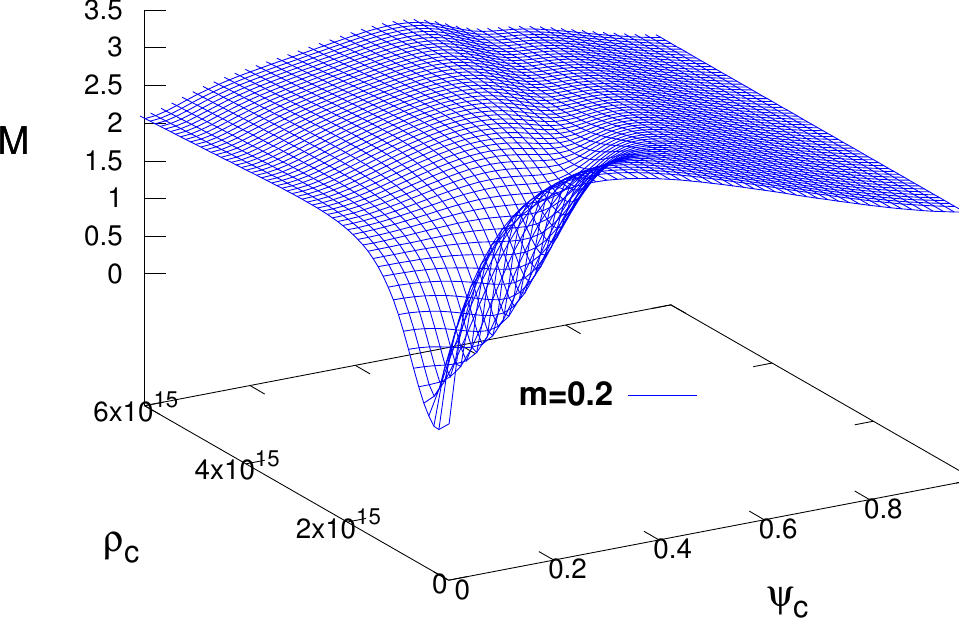}
	\includegraphics[width=0.48\textwidth]{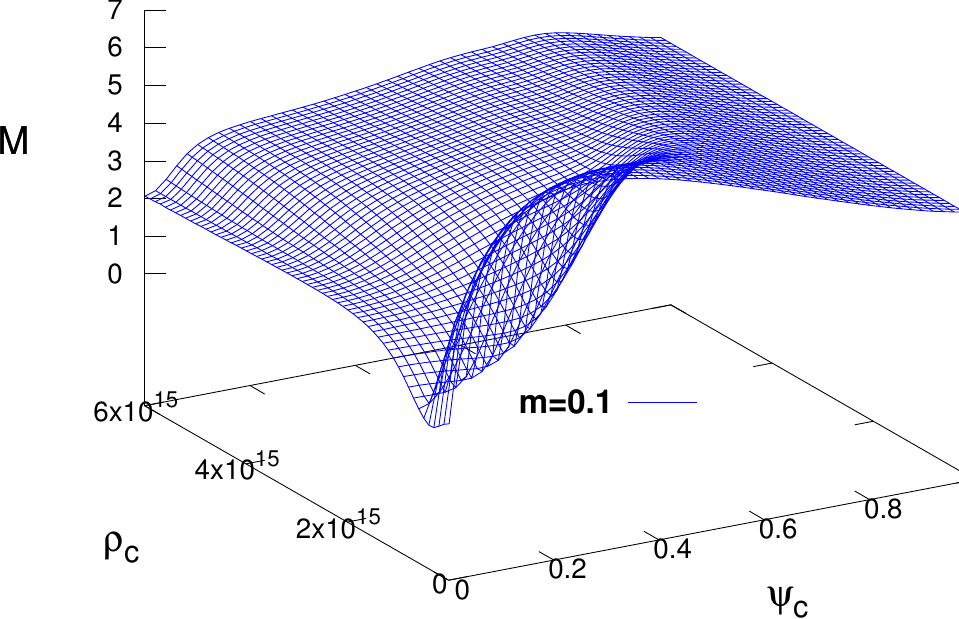}
	\caption{The mass of the compact object as a function of the central energy density and the central value of the scalar field for conformal factor $A(\varphi)= \exp(\frac{1}{2}\beta\psi^2)$, $\beta=-6$ and $\kappa=0$. Four figures with different values of the scalar field mass are shown.}
	\label{fig:3D_var_m}
\end{figure}

In Fig. \ref{fig:3D_var_m} we have plotted the mass of compact object as a function of ${\tilde \rho}_c$ and $\psi_c$. Graphs for several different  values of the scalar field mass $m$ are shown and $\kappa$ is fixed to zero for simplicity as a first step. For large $m$ the mass of the pure soliton part is small while it can considerably increase with the decrease of $m$. We have chosen some representative values of $m$ in agreement with the constraints above which also lead to pure soliton mass of the same order as the mass of the pure neutron star.  Naturally, for small ${\tilde \rho}_c$  the mass is dominated by the scalar field while for small $\psi_c$ -- by the baryon matter. In the limits when either ${\tilde \rho}_c=0$ or $\psi_c=0$ the solutions posses the well known behavior -- the mass of the pure soliton or the pure neutron star, respectively, increases, reaches a maximum and then decreases where the maximum of the mass is associated with a  change of stability. The stability of the mixed soliton-star configurations will be discussed below.

Naturally for small values of ${\tilde \rho}_c$ the soliton contribution dominates while for small $\psi_c$ the baryon contribution is more pronounced.  An interesting observation is that for large $\psi_c$ the baryon part is suppressed even for very large center energy densities. For a constant ${\tilde \rho}_c$ and large $m$, the mass $M$ rapidly decrease with the increase of $\psi_c$ while in the small $m$ regime, $M$ increases reaching the typical values for a pure soliton. A general observation for the studied range of parameters is that for $\psi_c$ above roughly $0.5$ the baryon contribution is almost negligible and the mass $M$ is nearly independent of ${\tilde \rho}_c$. As we will show, though, for such a range of parameters the solutions are unstable. 
  
\begin{figure}
	\includegraphics[width=0.48\textwidth]{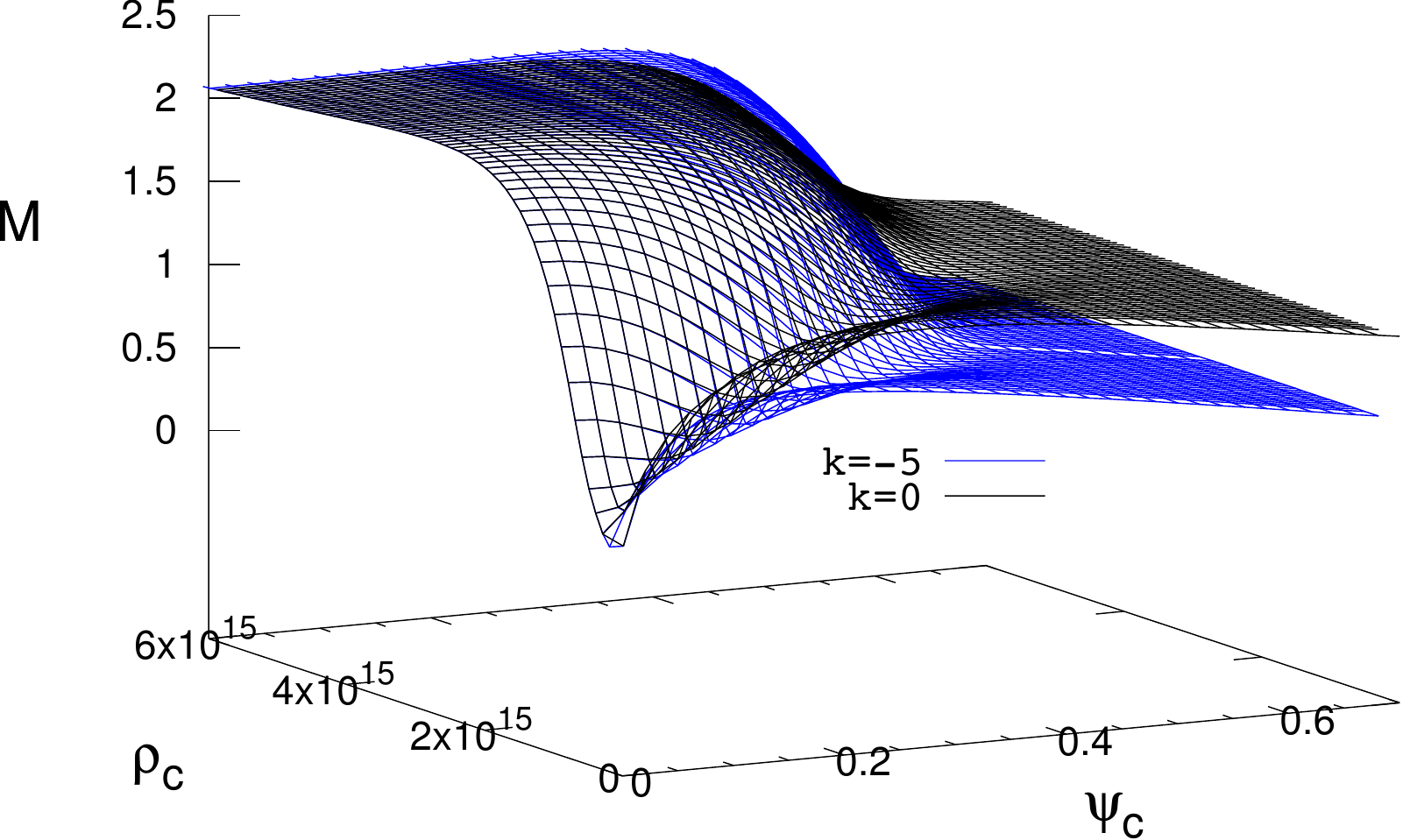}
	\includegraphics[width=0.48\textwidth]{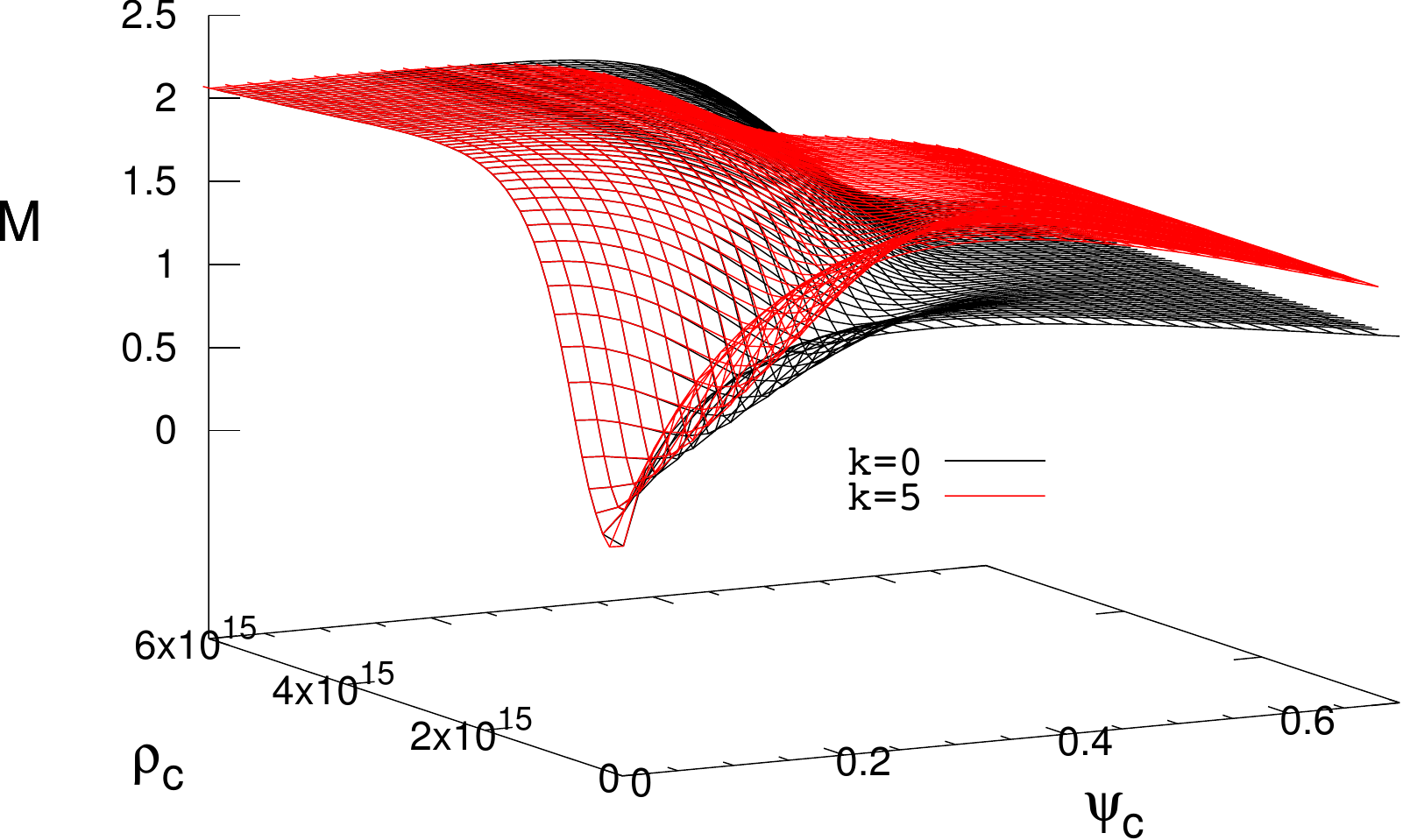}
	\caption{The mass of the compact object as a function of the central energy density and the central value of the scalar field  for conformal factor $A(\varphi)= \exp(\frac{1}{2}\beta\psi^2)$, $\beta=-6$, $m=0.5$ and different values of $\kappa$. In the \textit{left panel} the case of $\kappa=0$ and $\kappa=-5$ are compared while in the \textit{right panel} the cases of $\kappa=0$ and $\kappa=-5$ are examined. Two different graphs are shown in order to achieve a better visibility of the effect that changing $\kappa$ has on the results.}
	\label{fig:3D_var_kappa}
\end{figure}

\begin{figure}
	\includegraphics[width=0.48\textwidth]{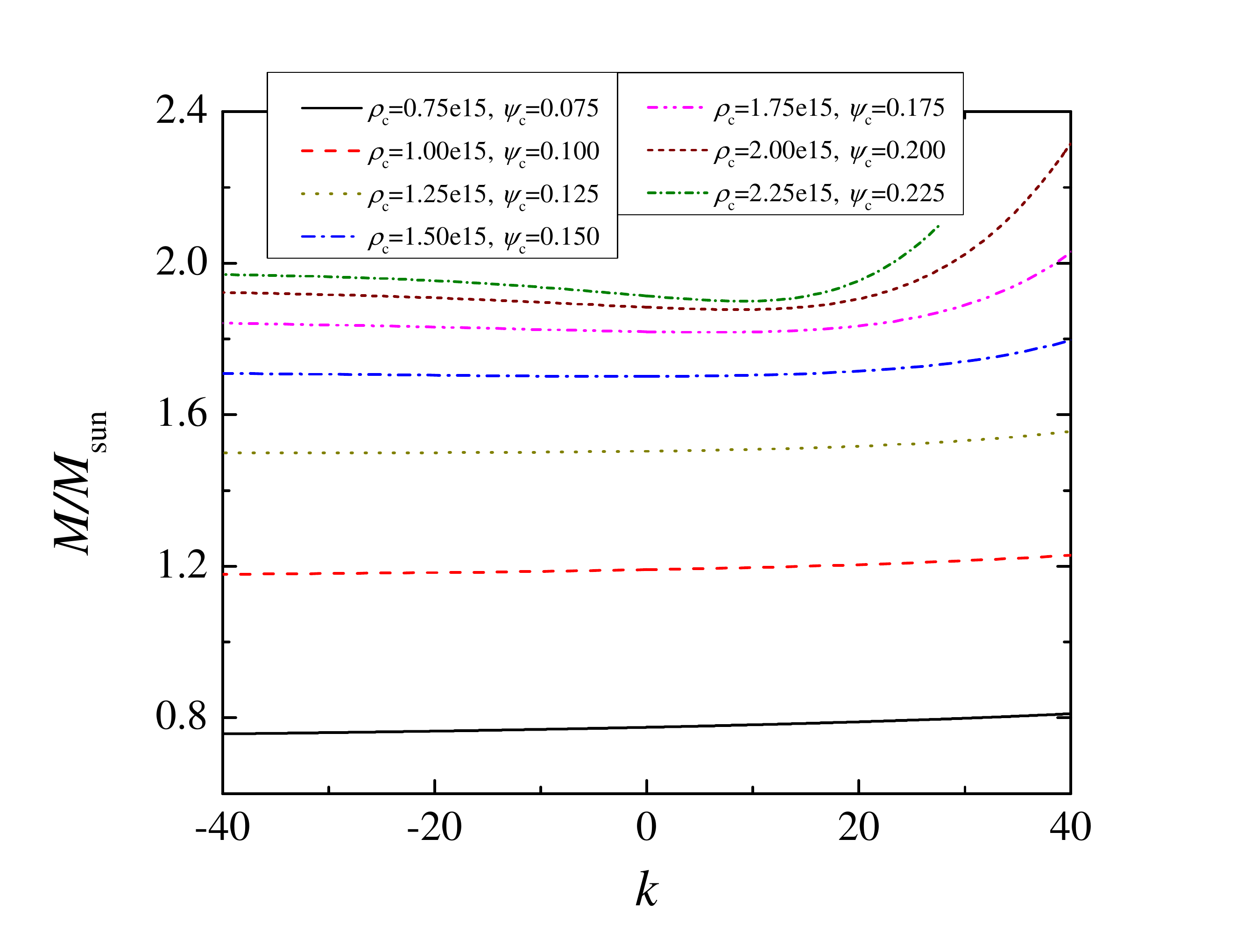}
	\includegraphics[width=0.48\textwidth]{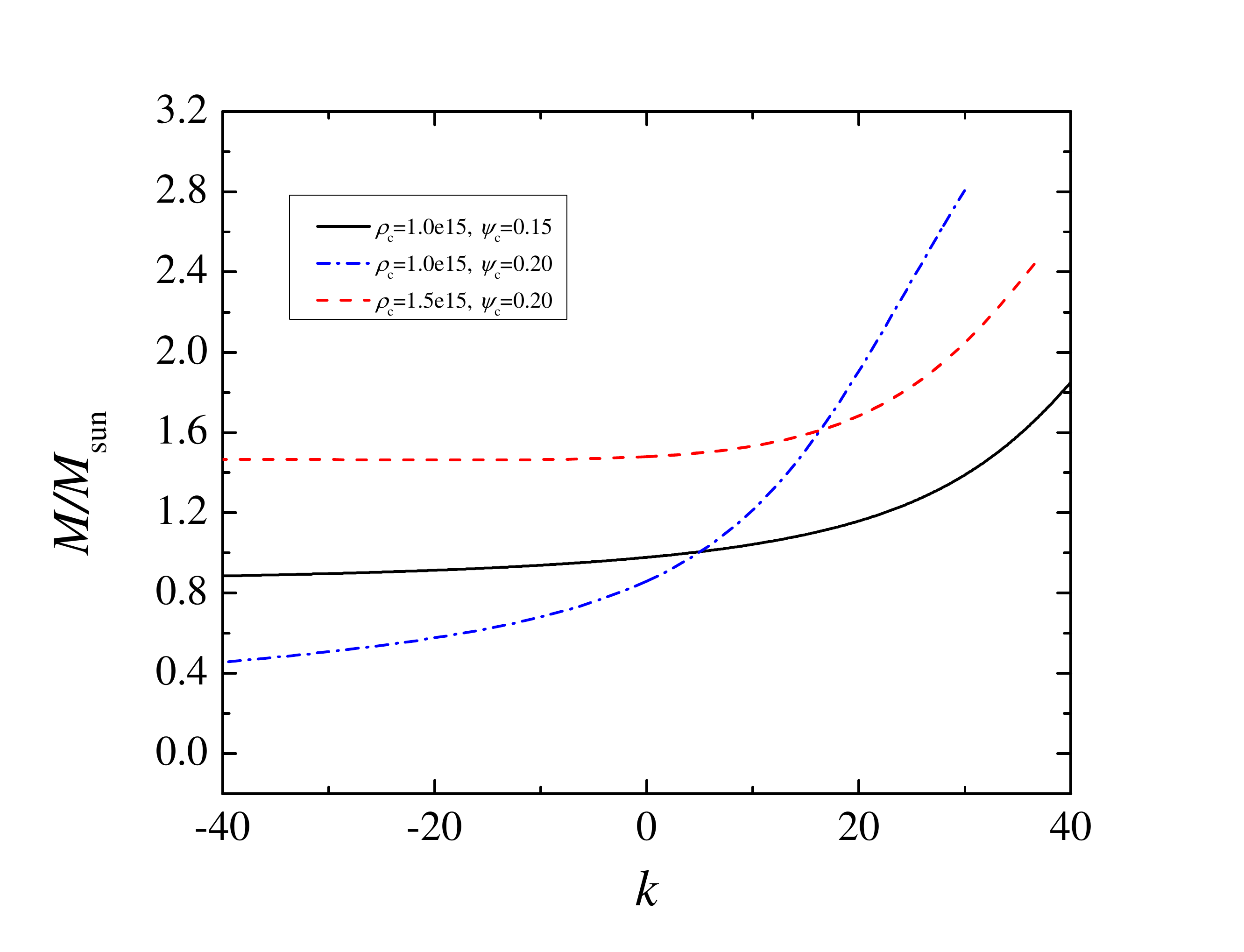}
	\caption{The dependence of the total mass of the mixed compact object as a function of the curvature constant $\kappa$ for different combinations of the ${\tilde \rho}_c$ and $\psi_c$, where we fix $\beta=-6$ and $m=0.5$. In the left panel combinations with comparable soliton and baryon contributions are plotted, while in the right panel the soliton part dominates over the baryon one and thus the $\kappa$ dependence is more pronounced.}
	\label{fig:2D_var_kappa}
\end{figure}

In Fig. \ref{fig:3D_var_kappa} we have examined the dependence of the compact object mass on the parameter $\kappa$. In the two panels the cases of $\kappa=5$ and $\kappa=-5$ are compared with the $\kappa=0$ results. As expected, for small $\psi_c$ or large ${\tilde \rho}_c$, the mass $M$ is practically unaltered by the change of $\kappa$. For large values of the scalar field the total mass of the object follows the behavior already studied in \cite{Yazadjiev_2019} -- for positive curvatures $\kappa$ the mass increases while for small $\kappa$ it decreases. This behavior can change for intermediate values of $\psi_c$ and ${\tilde \rho}_c$ and the mass slightly increases for $\kappa<0$ while it decreases for $\kappa>0$. As a matter of fact, as we will see below, this is exactly the region where a change of stability of the mixed configuration is observed. The dependence of the total mass $M$ on the parameter $\kappa$ can be observed as well in Fig. \ref{fig:2D_var_kappa} where the $M$ is plotted as a function of $\kappa$ for several different combinations of ${\tilde \rho}_c$ and $\psi_c$. As one can see in the left panel, for large enough ${\tilde \rho}_c$ compared to $\psi_c$, the effect of the target space metric curvature is pronounced only for very large $\kappa$. In the right panel  ${\tilde \rho}_c$ and $\psi_c$ are chosen such that the soliton and the baryon part have similar contribution which leads to stronger dependence of $M$ on the curvature $\kappa$. We have not plotted models where the soliton part strongly dominated over the baryon one because as we will see below these models are supposed to be unstable. 

\begin{figure}
	\includegraphics[width=0.48\textwidth]{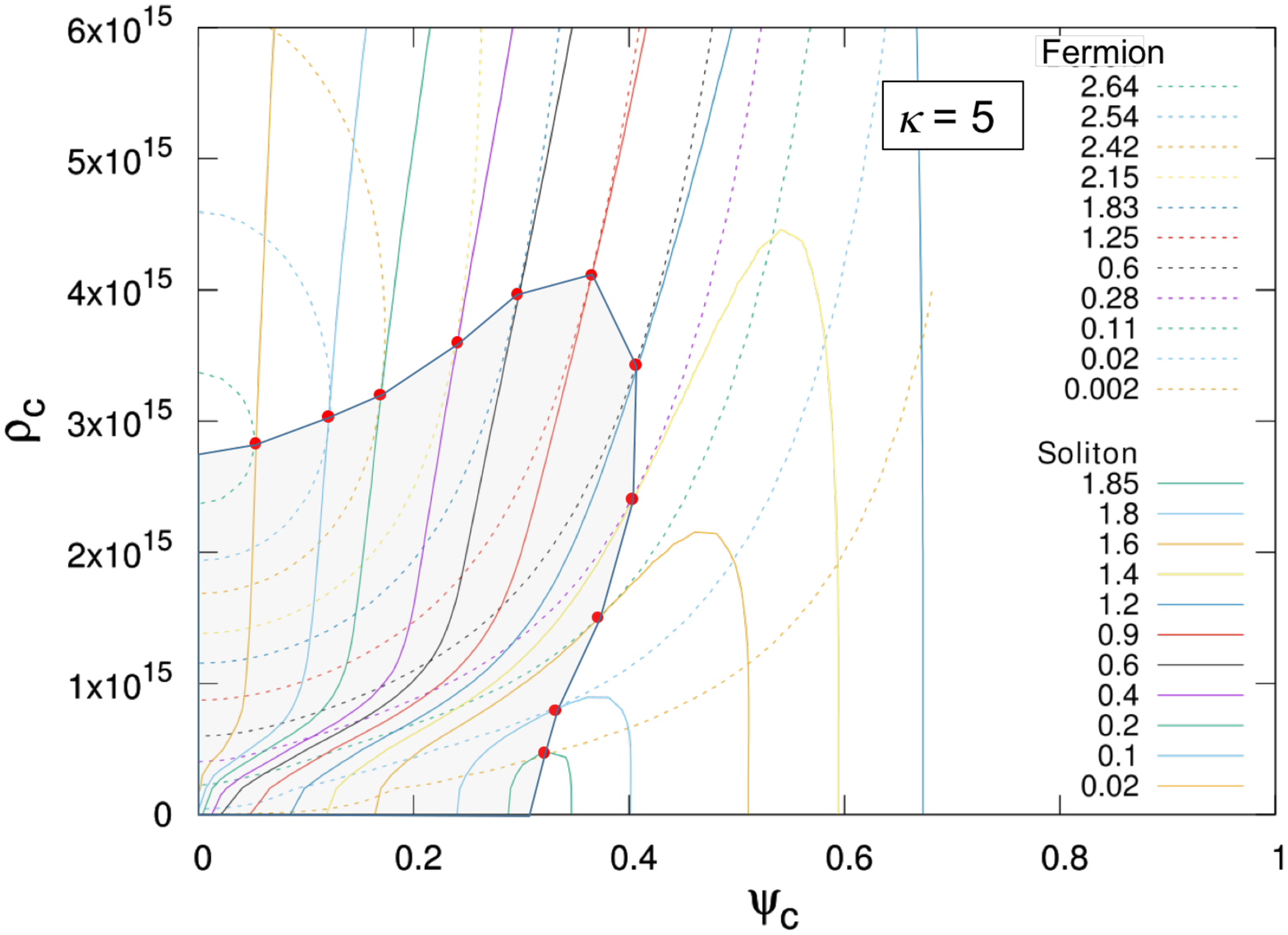}
	\includegraphics[width=0.48\textwidth]{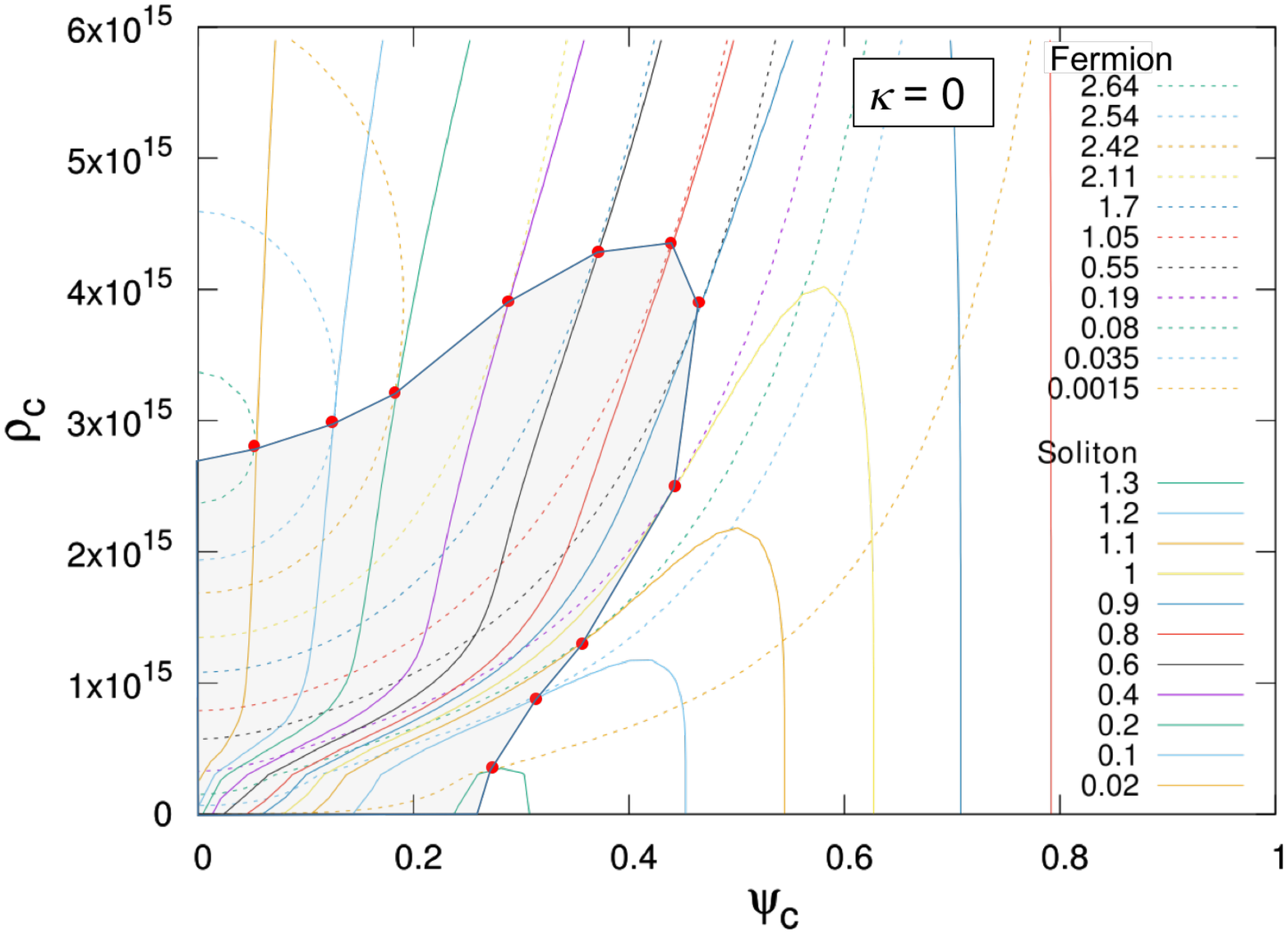}
	\includegraphics[width=0.48\textwidth]{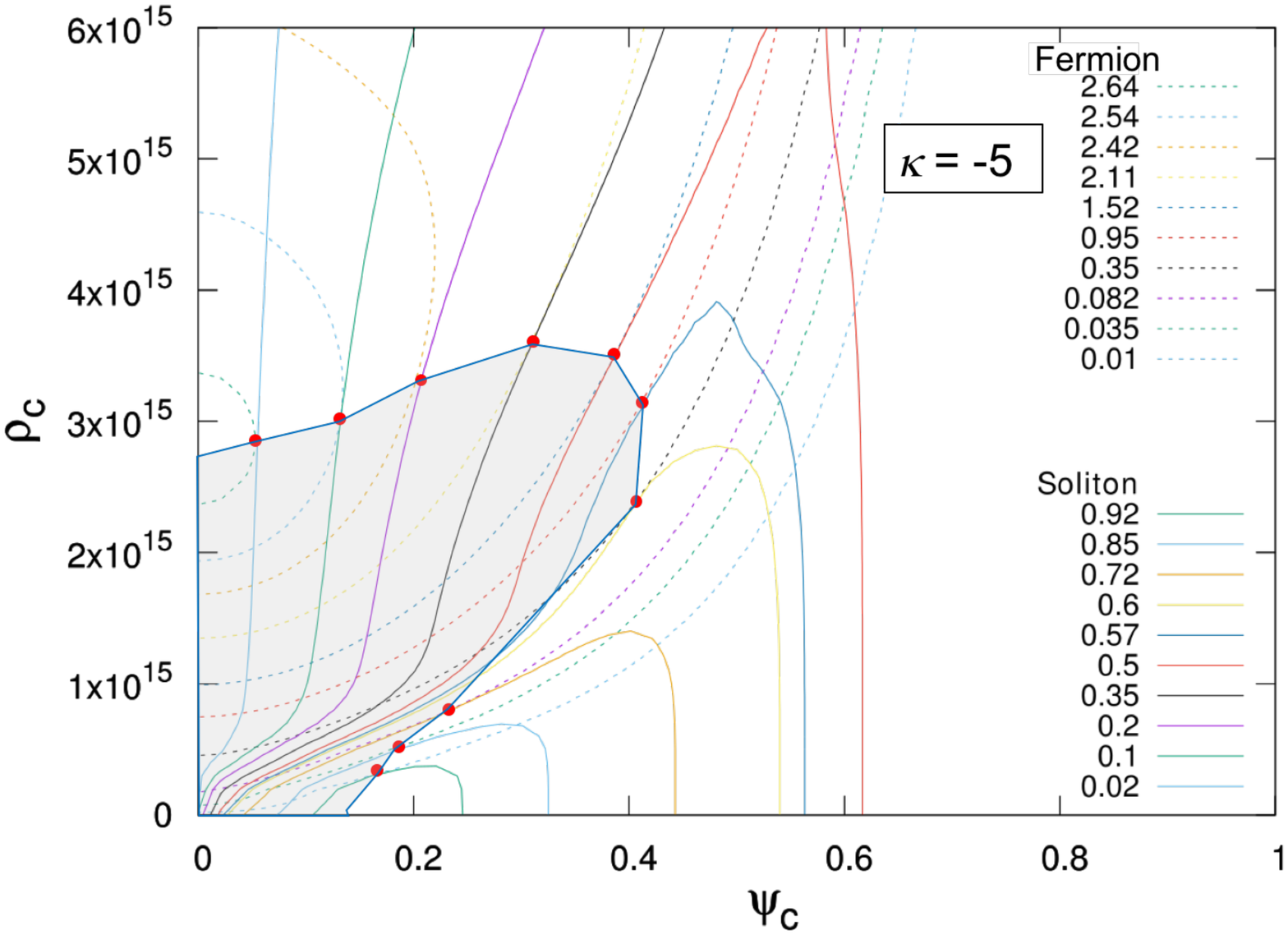}
	\caption{Plots of the stability region for $\gamma=1$, $m=0.5$ and different values of $\kappa$. The conformal factor is $A(\varphi)= \exp(\frac{1}{2}\beta\psi^2)$. Solid lines correspond to constant $Q$ while dashed lines correspond to constant baryon mass $M_0$. At the points where the two sets of lines touch a change of stability is observed and the shaded area denotes the stable region.}
	\label{fig:2D_contour}
\end{figure}

Let us now concentrate on the stability of the mixed soliton-star solutions. In the case of zero $\psi_c$ or ${\tilde \rho}_c$ the change of stability occurs at the maximum of the mass. In the case of mixed configurations  the situation with the stability is more complicated. Fortunately it  can be mathematically treated as the standard boson-fermion stars in General relativity \cite{Henriques_1990c}. The change of stability as well know is related to the static perturbations
$(\delta\psi_c,\delta{\tilde \rho}_c)$ leading from an equilibrium state $(\psi_c,{\tilde \rho}_c)$ to another equilibrium state $(\psi_c +\delta\psi_c, {\tilde \rho}_c+ \delta{\tilde \rho}_c )$ with  the same  $M$, $M_{0}$ and $Q$. Consequently, as shown in \cite{Henriques_1990c}, if there is a point  $P$
in the $(\psi_c, {\tilde \rho}_c)$ plane where the stability changes,  there must be a direction ${\vec n}$ at $P$ such that the directional derivatives satisfy 

\begin{eqnarray}\label{SCL}
\frac{\partial M}{\partial{\vec n}}|_P=\frac{\partial M_{0}}{\partial{\vec n}}|_P=\frac{\partial Q}{\partial{\vec n}}|_P=0.
\end{eqnarray} 
Therefore the locus of points in the plane $(\psi_c,{\tilde \rho}_c)$ where the stability changes is the locus of points satisfying 
(\ref{SCL}). In order to determine the dimension of the locus we have to take into account the following relation among $\delta M$, $\delta M_{0}$ and $\delta Q$, namely 
\begin{eqnarray}\label{RMQ}
\delta M= \frac{\mu}{m_b}\delta M_{0} + \frac{\omega}{4\pi m}\delta Q
\end{eqnarray}
with $\mu$ being the red-shifted chemical potential of the baryons. This relation can be derived using a procedure similar to 
those described in \cite{Jetzer_1992} and \cite{Yazadjiev_1999}. With the ralation (\ref{RMQ}) in mind we can conclude that the loci where the stability changes 
are lines in the $(\psi_c,{\tilde \rho}_c)$ plane. In practice, as for the general relativistic boson-fermion stars,      
in the mixed  soliton-star case the following procedure can be performed in order to determine the stability region \cite{Henriques_1990c}.  One draws in a two dimensional $\{\psi_c, {\tilde \rho}_c\}$ plot, lines of constant baryon rest mass $M_0$ and constant conserved charge $Q$. At the points where the two sets of lines intersect a change of stability is observed. Such plots, for the values of $\kappa$ examined in Fig. \ref{fig:3D_var_kappa}, is depicted in Fig. \ref{fig:2D_contour}. As one can see the region of stability spans to larger values of $\psi_c$ and ${\tilde \rho}_c$ compared to the pure soliton and pure neutron star case. 

The stability region increases with the increase of $\kappa$ and decreases for $\kappa<0$ . The qualitative behavior is qualitatively similar, though, for different values of $\kappa$.

\subsection{Theory with $A(\varphi)= \exp(\frac{1}{4}\gamma\psi^4)$}
\begin{figure}
	\includegraphics[width=0.48\textwidth]{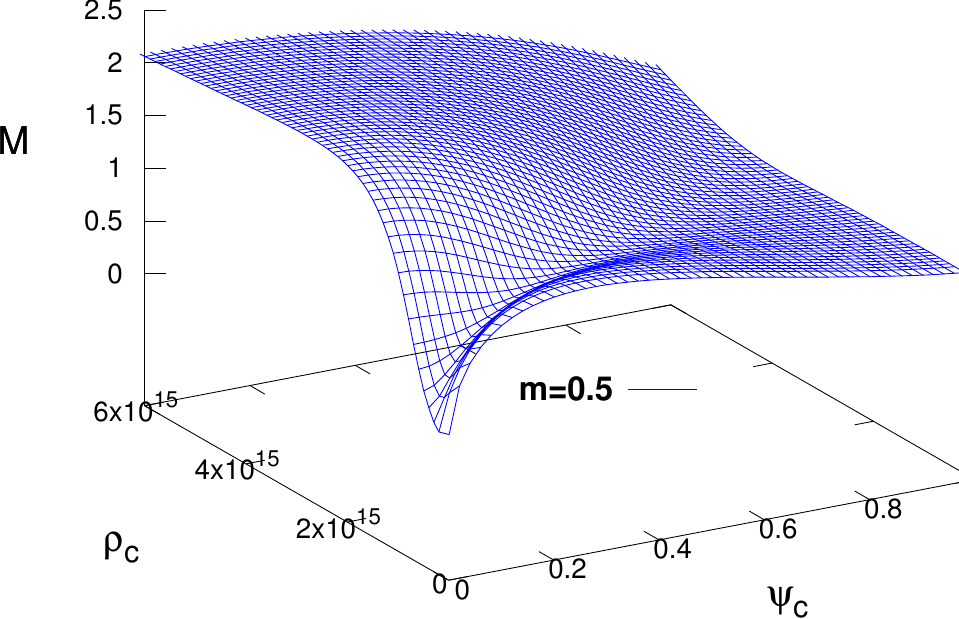}
	\includegraphics[width=0.48\textwidth]{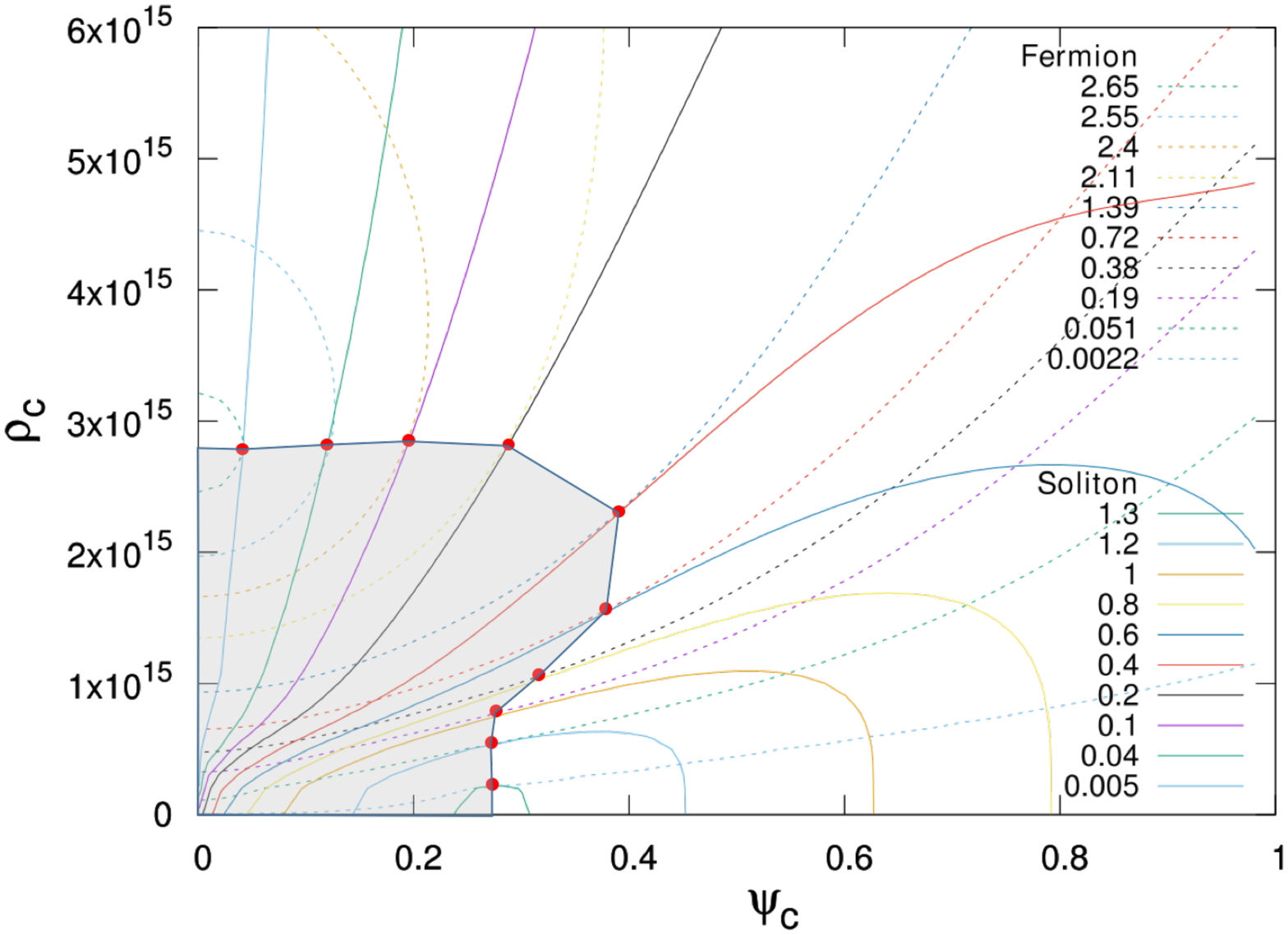}
	\caption{(left panel) The mass of the compact object as a function of the central energy density and the central value of the scalar field for  the conformal factor $A(\varphi)= \exp(\frac{1}{4}\gamma\psi^4)$, where $\gamma=1$, $m=0.5$ and $\kappa=0$. (right panel) The stability region for the same values of the parameters.}
	\label{fig:psi4_3D}
\end{figure}

In this section we will concentrate on a different conformal factor, namely
\begin{equation}
A(\varphi)= \exp(\frac{1}{4}\gamma\psi^4).
\end{equation}
This  scalar-tensor theory with $\gamma>0$ is particularly interesting because, as discussed in \cite{Yazadjiev_2019}  it belongs to a class of theories for which the scalar fields are not excited neither in neutron stars (with ${\tilde \rho \ge 3 {\tilde p}}$)  nor in the weak field limit \cite{Yazadjiev_2019}. For such theories no constraints can be imposed in practice by the current observations. However, the scalar fields are do excited in the tensor-multi-scalar solitons. In Fig. \ref{fig:psi4_3D} we have plotted a 3D graph of the total mass of the mixed soliton-fermion configuration as a function of ${\tilde \rho}_c$ and $\psi_c$ for $\gamma=1$, $m=0.5$ and $\kappa=0$. As one can see the qualitative behavior is similar to the one observed in Fig. \ref{fig:3D_var_m} for the conformal factor $A(\varphi)= \exp(\frac{1}{2}\beta\psi^2)$ with the following important difference. For a fixed ${\tilde \rho}_c$, the mass decreases smoothly as the scalar field increases  for the $\exp(\frac{1}{4}\gamma\psi^4)$ conformal factor compared to  the $\exp(\frac{1}{2}\beta\psi^2)$ case where the decrease is much sharper. As a result, even for large central values of the scalar field the baryon part is not completely suppressed. If we look at the right panel of Fig. \ref{fig:psi4_3D}, though, the stability region is actually smaller than the one observed for $A(\varphi)=\exp(\frac{1}{2}\beta\psi^2)$. Thus it is evident that not only the introduction of a curvature of the target space metric has non-negligible effect and modifies the compact object solutions compared to the pure boson-fermion stars, but also the conformal factor $A(\varphi)$ has significant effect on the structure of the solutions.

We should note that the above conclusions for the differences between the two conformal factors are made using specific values of the parameters $\beta$ and $\gamma$. Clearly, the structure of the solutions depends on these parameters as well, but since we have a many parameter family of solutions we have decided to fix them. The discussion above should be viewed more as a proof for the large effect the conformal factor has on the soliton-baryonic compact objects and the richness of the choices of such a factor.

\begin{figure}
	\includegraphics[width=0.48\textwidth]{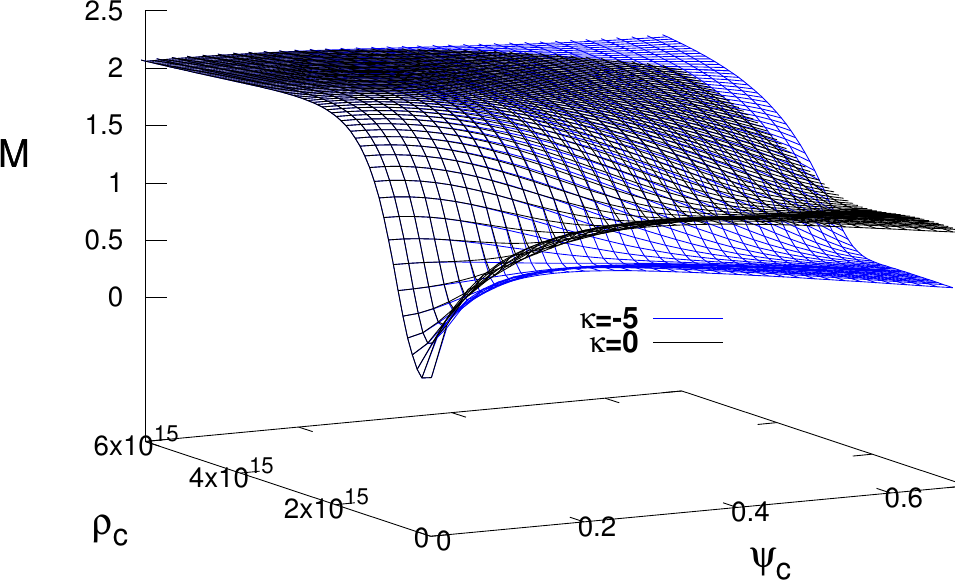}
	\includegraphics[width=0.48\textwidth]{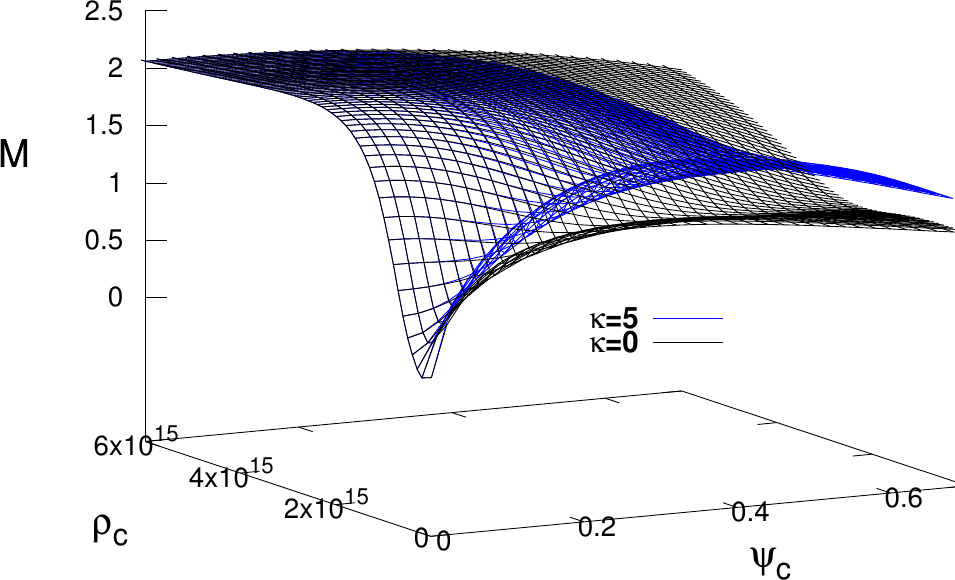}
	\caption{The mass of the compact object as a function of the central energy density and the central value of the scalar field for the conformal factor $A(\varphi)= \exp(\frac{1}{4}\gamma\psi^4)$,  $\gamma=1$, $m=0.5$ and different values of $\kappa$ are shown. In the \textit{left panel} the case of $\kappa=0$ and $\kappa=-5$ are compared while in the \textit{right panel} the cases of $\kappa=0$ and $\kappa=-5$ are examined. Two different graphs are shown in order to achieve a better visibility of the effect that changing $\kappa$ has on the results.}
	\label{fig:3D_var_kappa_psi4}
\end{figure}

\section{Conclusion}
In the present paper we examined mixed configurations of tenor-multi-scalar solitons and relativistic (neutron) stars within a special class of massive  tensor-multi-scalar theories of gravity whose target space metric admits Killing field(s) with a periodic flow. These objects are similar to the pure boson-fermion compact objects, but they can have much richer phenomenology. The reason is that we can choose different forms of the target space metric  while in the pure boson-fermion case the  target space metric is just the flat one with curvature parameter $\kappa=0$. Moreover, what makes the difference with the standard boson-fermion stars even more drastic is the highly nonstandard coupling between the effective boson $\psi$ field and the matter -- the effective boson field is sourced by the trace of the perfect fluid  energy-momentum tensor in the Einstein frame and the coupling  between $\psi$  and the perfect fluid  depends  very strongly on the particular scalar-tensor theory. 

We have a many parameter family of solutions that is determined on one hand by the central value of the baryonic matter energy density and central values of the scalar field, and on the other hand by the curvature parameter in the target space metric $\kappa$, the parameter in the conformal factor and the scalar field mass $m$. We have explored a variety of combinations of the above parameters concentrating mainly on the effect $\kappa$ and $m$ have on the properties of the solutions. The increase (decrease) of the scalar field mass in general leads to smaller (larger) mass of the soliton part. In the pure soliton case, the mass of the compact object increases for positive $\kappa$ and decreases for negative. This behavior is no longer monotonic with the inclusion of a baryonic part and there are some region of the $\{{\tilde \rho}_c, \psi_c\}$ parameter space, where the total mass of the mixed object decreases (increases) for positive (negative) $\kappa$.

A general observations is that for large central values of the scalar field $\psi_c$ the mixed configuration is dominated by the soliton contribution, even for central energy densities of the baryonic matter ${\tilde \rho}_c$ close to the maximum ones allowed for pure neutron stars. Thus for small $m$ the pure soliton part has larger mass than the pure baryonic part thus leading to an increase of the total mass with the increase of $\psi_c$ for all of the studied ${\tilde \rho}_c$. On the contrary, for larger $m$ the mass of the pure solitons is much smaller than the pure neutron star mass and for large $\psi_c$ the baryonic contribution is suppressed. We should note, though, that his normally happen in the region where the mixed configurations are supposed to be unstable. The stability studies showed as well that the stability region for mixed soliton-baryonic stars spans to larger values of $\psi_c$ and ${\tilde \rho}_c$ compared to the pure boson or the pure fermion configurations.  

We have examined two different forms of the conformal factor. The one given by $A(\varphi)= \exp(\frac{1}{2}\beta\psi^2)$ is motivated by studies of scalarized neutron stars, while the second one $A(\varphi)= \exp(\frac{1}{4}\gamma\psi^4)$ is particularly interesting because in this case the scalar fields are not excited neither in neutron stars nor in the weak field limit and therefore no observational constraints can be imposed from the binary pulsar observations. Even though the mixed soliton-baryonic compact objects share similar properties in both cases, there are important differences which demonstrate that the richness of the solutions compared to the pure boson-fermion case where $A(\varphi)=1$.

\section*{Acknowledgements}
DD would like to thank the European Social Fund, the Ministry of Science, Research and the Arts Baden-Wurttemberg for the support. DD is indebted to the Baden-Wurttemberg Stiftung for the financial support of this research project by the Eliteprogramme for Postdocs. DD acknowledge financial support via an Emmy Noether Research Group funded by the German Research Foundation (DFG) under grant
no. DO 1771/1-1.
SY acknowledges financial support by the Bulgarian NSF Grant DCOST 01/6. Networking support by the COST Actions  CA16104 and CA16214 is also gratefully acknowledged.



\end{document}